\documentclass[11pt]{article}

\usepackage{amsfonts,amssymb,chet,dsfont,graphicx,mathrsfs,pgfplots,tikz}
\usetikzlibrary{pgfplots.fillbetween}
\usetikzlibrary{external}
\newcommand{\Pochhammer}[2]{\left({#1}\right)_{#2}}

\title{A Mellin Transform Approach\\to Rephasing Invariants}

\author{Jean-Fran\c{c}ois Fortin\email{jean-francois.fortin@phy.ulaval.ca}, Nicolas Giasson\email{nicolas.giasson.1@ulaval.ca}, Luc Marleau\email{luc.marleau@phy.ulaval.ca} and Jasmine Pelletier-Dumont\email{jasmine.pelletier-dumont.1@ulaval.ca}}

\affiliation{
D\'epartement de Physique, de G\'enie Physique et d'Optique,\\Universit\'e Laval, Qu\'ebec, QC G1V 0A6, Canada
}

\abstract{%
In the low-energy effective theory of neutrinos, the Haar measure for unitary matrices is very likely to give rise to the observed PMNS matrix.  Assuming the Haar measure, we determine the probability density functions for all quadratic, quartic Majorana, and quartic Dirac rephasing invariants for an arbitrary number of neutrino generations.  We show that for a fixed number of neutrinos, all rephasing invariants of the same type have the same probability density function under the Haar measure.  We then compute the moments of the rephasing invariants to determine, with the help of the Mellin transform, the three probability density functions.  We finally investigate the physical implications of our results in function of the number of neutrinos.
}

\date{April 2020} 

\begin{document}

\maketitle



\section{Introduction}\label{SecIntro}

In flavor physics, the passage from gauge eigenstates to mass eigenstates encodes flavor mixing.  This mixing is encapsulated in the Cabibbo-Kobayashi-Maskawa (CKM) matrix for the quark sector.  In the Standard Model of particle physics, there is no equivalent mixing for the lepton sector.  However, the Standard Model must be extended to take into account neutrino oscillations \cite{pdg,Esteban:2016qun}, and that extension allows for mixing in the lepton sector.  In the low-energy effective theory of neutrinos, this is encoded in the Pontecorvo-Maki-Nakagawa-Sakata (PMNS) matrix for the lepton sector.

The CKM and PMNS mixing matrices, which are unitary matrices, can be redefined by phase rotations of the quark and lepton fields, respectively.  Since physical observables must be invariant under these field redefinitions, only some functions of the mixing matrix elements can be measured explicitly.  The simplest way to proceed is to write physical observables in terms of the so-called rephasing invariants of the mixing matrices \cite{Jarlskog:1985ht,Greenberg:1985mr,Dunietz:1985uy}.  As their name implies, rephasing invariants do not change under field redefinitions.  The most celebrated rephasing invariant is the Jarlskog invariant \cite{Jarlskog:1985ht} associated to the CP-violating Dirac phase of the CKM matrix.

Flavor physics is notoriously hard.  Experimental data show that the CKM matrix is hierarchical while the PMNS matrix is rather random, with a preference for near-maximal mixing.  It is very difficult to come up with a convincing theoretical story behind the patterns observed in the mixing matrices.  One possible path forward is to study the mixing matrices statistically.  Indeed, it is possible to determine how likely it is to draw at random a unitary matrix resembling the CKM matrix or the PMNS matrix from a given probability density function (PDF).  If that probability is large, then the mixing matrix is likely to originate from the associated PDF, and the average values of the different rephasing invariants under that PDF can be compared with the observed experimental values, leading to predictions for the unknown ones.

For the quark sector, the CP-violating Jarlskog invariant mentioned above was studied statistically in \cite{Gibbons:2008su,Dunkl:2009sn}.  Assuming the Haar measure, which is the most natural measure on the space of unitary matrices, the PDF for the Jarlskog invariant was computed analytically in \cite{Dunkl:2009sn}.  Considering that the observed Jarlskog invariant is $|y_\text{exp}^D|_\text{CKM}=(3.04_{-0.20}^{+0.21})\times10^{-5}$ \cite{pdg} and the probability of obtaining it from the PDF associated to the Haar measure is very small $P\{|y^D|\leq|y_\text{exp}^D|_\text{CKM}\}\approx0.08\%$, it was shown in \cite{Dunkl:2009sn} that the CKM matrix should not be seen as being a generic unitary matrix drawn randomly from the PDF associated to the Haar measure.

For the lepton sector, an equivalent analysis was performed in \cite{Fortin:2018etr}.  It was shown there that under the Haar measure, the probability of generating a unitary matrix with the observed quartic Dirac rephasing invariant $|y_\text{exp}^D|_\text{PMNS}=0.032_{-0.005}^{+0.005}$ (see for example \cite{Abe:2017uxa,Abe:2017vif}) was quite large, $P\{|y^D|\leq|y_\text{exp}^D|_\text{PMNS}\}\approx60\%$.  Allowing for the possibility that neutrinos are Majorana, the same was true for the quartic Majorana rephasing invariants.  Hence \cite{Fortin:2018etr} concluded that the statistical hypothesis that the PMNS matrix arises randomly from the PDF associated to the Haar measure was highly likely, contrary to the CKM matrix.  Moreover, \cite{Fortin:2018etr} showed that the average value of the quartic Dirac rephasing invariant $\langle|y^D|\rangle_\text{PMNS}=\pi/105\approx0.030$ was in striking agreement with the observed value.  Since CP violation is more important in the lepton sector, the statistical analysis of \cite{Fortin:2018etr} thus suggests that the baryon asymmetry of the Universe could originate from leptogenesis.

Although the Haar measure is the most natural measure for unitary matrices, there is a plausible theoretical story behind its origin, namely the anarchy principle \cite{Hall:1999sn,deGouvea:2003xe,Heeck:2012fw,deGouvea:2012ac,Bai:2012zn,Lu:2014cla,Babu:2016aro,Long:2017dru,Haba:2000be,Espinosa:2003qz,Fortin:2016zyf,Fortin:2017iiw}.  The anarchy principle states that the light neutrino mass matrix parameters originate from the seesaw mechanism and that the high-energy mass matrices are generated randomly from the appropriate Gaussian ensembles.  The low-energy neutrino parameters are thus derived from these randomly-generated high-energy parameters, leading to specific ensembles for the low-energy parameters \cite{Fortin:2016zyf,Fortin:2017iiw,Fortin:2018etr}.  It was then proven there that the PDF for arbitrary neutrino numbers factorizes into a PDF for the light neutrino mass eigenvalues and a PDF for the mixing angles and phases of the PMNS matrix.  The former is given by a complicated multidimensional integral while the latter is simply the Haar measure (independently of the seesaw mechanism, as foreseen on physical grounds in \cite{Haba:2000be}).  The factorization into two independent PDFs for the light neutrino masses and mixing parameters leads to physical implications that are independent between the masses and the PMNS matrix.  For the masses, it was shown that the preferred seesaw mechanism is of type I-III while the preferred mass splitting is in agreement with the normal hierarchy.

The PDFs for the PMNS (or, for that matter, the CKM) rephasing invariants associated to the $U(N)$ Haar measure for neutrino numbers $N=2$ and $N=3$ were obtained in \cite{Fortin:2018etr} based on the work of \cite{Dunkl:2009sn}.  The technique employed there was built on the knowledge of the moments being expressed as products of beta-distributed random variables.  Although explicit, it was unclear how complicated the PDFs would become for larger neutrino numbers which could be of interest for extensions of the Standard Model with sterile neutrinos.  In this paper, we introduce another technique relying on the knowledge of the moments and the Mellin transform.  This method leads to direct expressions for all rephasing invariant PDFs for arbitrary neutrino numbers in terms of Meijer $G$-functions.  In this unified theoretical formalism, we will demonstrate that all rephasing invariants of the same type (\textit{i.e.} quadratic, quartic Majorana, and quartic Dirac) have the same PDF.  In function of the number of neutrinos $N$, we will also argue that the anarchy principle, and more generally the Haar measure, prefers three neutrino flavors.

This paper is organized as follows: Section \ref{SecReview} discusses quadratic, quartic Majorana, and quartic Dirac rephasing invariants.  The Haar measure is then introduced and some of its properties are demonstrated.  A convenient parametrization for unitary matrices is also described.  The equality of the PDFs for rephasing invariants of the same type is then proven with the help of permutation matrices.  In Section \ref{SecMellin} the Mellin transform approach to PDFs is discussed in all generality and some preliminary results on Meijer $G$-functions are given.  In Section \ref{SecPDF}, the PDFs for the three types of rephasing invariants are computed in function of the neutrino number and the results are expressed in terms of the Meijer $G$-functions for the quartic rephasing invariants.  Section \ref{SecDisc} presents a discussion of the analytic results, with comparisons to numerical results, an analysis of the behavior of the PDFs around the origin, and an analysis of the average values in function of the neutrino number.  For the latter, it is shown that the observed experimental values prefer three neutrino flavors.  Finally, Section \ref{SecConc} presents our conclusion.


\section{Review}\label{SecReview}

In this section we discuss the quadratic, quartic Majorana, and quartic Dirac rephasing invariants.  After reviewing the Haar measure, we demonstrate that all rephasing invariants of the same type (quadratic, quartic Majorana, quartic Dirac) have the same PDFs with respect to the Haar measure.  Hence, there are only three distinct PDFs to consider for any neutrino number $N$.


\subsection{Rephasing Invariants}\label{subsec:RephasingInv}

As stated in the introduction, basis independence implies that the proper physical observables obtained from the PMNS matrix must be invariant under phase rotations of the fields.  These physical observables are the rephasing invariants \cite{Jarlskog:1985ht,Greenberg:1985mr,Dunietz:1985uy}.  For the unitary matrix $U$, the quadratic $y_{ij}$, quartic Majorana $y_j^M$, and quartic Dirac $y_{ij}^D$ rephasing invariants are given by \cite{Jenkins:2007ip}
\eqn{
\begin{gathered}
x_{ij}=|U_{ij}|^2,\\
y_j^M=\text{Im}(U_{i_0j}U_{i_0j}U_{i_0j_0}^*U_{i_0j_0}^*),\\
y_{ij}^D=\text{Im}(U_{i_0j_0}U_{ij}U_{i_0j}^*U_{ij_0}^*),
\end{gathered}
}[EqRI]
respectively.  Here, the values $i_0$ and $j_0$ are fixed arbitrarily and the indices $i$ and $j$ labeling the different rephasing invariants are such that $i,j\neq i_0,j_0$.  To reach a set of independent rephasing invariants, other constraints must be imposed on the ranges of $i$ and $j$ \cite{Jenkins:2007ip}.  However, this observation is of no consequence since all rephasing invariants of the same type have the same PDFs as shown below.

Since the rephasing invariants \eqref{EqRI} are bounded as
\eqn{0\leq x_{ij}\leq1,\qquad-\frac{1}{4}\leq y_j^M\leq\frac{1}{4},\qquad-\frac{1}{6\sqrt{3}}\leq y_{ij}^D\leq\frac{1}{6\sqrt{3}},}[EqRIRange]
for future convenience it is of interest to rescale them in the following way,
\eqn{x_{ij},\qquad x_j^M=16|y_j^M|^2,\qquad x_{ij}^D=108|y_{ij}^D|^2.}[EqRIscaled]
Hence the three types of rescaled rephasing invariants $x$ are bounded on the interval $[0,1]$.  We note here that the rescaling \eqref{EqRIscaled} is motivated in parts by the fact that the odd moments of the quartic rephasing invariants under the Haar measure vanish.

Before proving that there are only three independent PDFs (one per type of rephasing invariants), we now focus on the Haar measure and discuss some of its properties.


\subsection{Haar Measure}

The Haar measure for the $N\times N$ unitary matrix $U$ is obtained straightforwardly by taking the wedge product of each independent elements of the matrix $U^\dagger dU$,\footnote{Although $U^\dagger dU$ is a matrix, we use the same notation for the measure.  The meaning should be clear from the context.} which arises naturally from singular value decomposition \cite{muirhead2009aspects}.  By definition, the Haar measure is both left- and right-invariant, \textit{i.e.} it satisfies $U^\dagger dU\to U^\dagger dU$ when $U\to LUR$ for $L$ and $R$ constant unitary matrices.  This property is easily proven since $U^\dagger dU\to R^\dagger U^\dagger dUR$ and the wedge product leads to
\eqn{(U^\dagger dU)\equiv\bigwedge_{1\leq i\leq j\leq N}(U^\dagger dU)_{ij}\to(R^\dagger U^\dagger dUR)=p(R)(U^\dagger dU),}
where $p(R)$ is a polynomial in $R$.  A simple computation shows that for $R=R_2R_1$, we must have $p(R_2R_1)=p(R_1)p(R_2)$, therefore the polynomial $p(R)$ must be a positive power of the determinant.  Clearly, since the Jacobian of any tranformation must be real, the Jacobian of the transformation $U\to LUR$ must be given by the norm of a positive power of the determinant, \textit{i.e.} $p(R)=|\det R|^k$ for some positive number $k$.  Hence, considering that $R$ is unitary, $p(R)=1$ irrespective of the value of $k$ and the Haar measure is both left- and right-invariant, as stated previously.

For future convenience, we now introduce a specific parametrization for unitary matrices based on \cite{1751-8121-43-38-385306,:/content/aip/journal/jmp/53/1/10.1063/1.3672064}.  In this parametrization, an $N\times N$ unitary matrix $U$ is expressed as
\eqn{U=\prod_{1\leq j<k\leq N}\exp(i\phi_{jk}P_k)\exp(i\theta_{jk}\Sigma_{jk})\prod_{1\leq j\leq N}\exp(i\varphi_jP_j),}[EqU]
where the matrices $P_j$ and $\Sigma_{jk}$ are given explicitly by
\eqn{(P_j)_{ik}=\delta_{ji}\delta_{jk},\qquad(\Sigma_{jk})_{i\ell}=-i\delta_{ji}\delta_{k\ell}+i\delta_{j\ell}\delta_{ki}.}
Here, the $N(N-1)/2$ mixing angles $\theta_{jk}$, the $N(N-1)/2$ phases $\phi_{jk}$, and the $N$ phases $\varphi_j$ are restricted to the intervals
\eqn{\theta_{jk}\in[0,\pi/2),\qquad\phi_{jk}\in[0,2\pi),\qquad\varphi_j\in[0,2\pi),}
respectively [implying the ranges \eqref{EqRIRange}].  Finally, the Haar measure in the parametrization \eqref{EqU} is given by
\eqn{U^\dagger dU=\prod_{1\leq i<j\leq N}\sin(\theta_{ij})[\cos(\theta_{ij})]^{2(j-i)-1}d\phi_{ij}d\theta_{ij}\prod_{1\leq i\leq N}d\varphi_i,}[EqHaar]
and depends non-trivially only on the mixing angles.

In the context of the PMNS matrix and the rephasing invariants \cite{Jenkins:2007ip}, the phases $\varphi_i$ are not the unphysical phases that can be absorbed by redefinitions of the fields.  Therefore, the usual CP-violating Majorana and Dirac phases are complicated functions of the phases $\varphi_i$ and the remaining phases $\phi_{ij}$.


\subsection{Equality of PDFs}

We now want to prove that all rephasing invariants of the same type have the same PDF.  To proceed, we focus on the moments of the rescaled rephasing invariants \eqref{EqRIscaled}, given by
\eqn{
\begin{gathered}
\left\langle(x_{ij})^{s-1}\right\rangle=\frac{1}{\text{Vol}\left(\mathcal{V}_N^2\right)}\int_{U\in\mathcal{V}_N^2}U^\dagger dU\,(x_{ij})^{s-1},\\
\left\langle(x_j^M)^{s-1}\right\rangle=\frac{1}{\text{Vol}\left(\mathcal{V}_N^2\right)}\int_{U\in\mathcal{V}_N^2}U^\dagger dU\,(x_j^M)^{s-1},\\
\left\langle(x_{ij}^D)^{s-1}\right\rangle=\frac{1}{\text{Vol}\left(\mathcal{V}_N^2\right)}\int_{U\in\mathcal{V}_N^2}U^\dagger dU\,(x_{ij}^D)^{s-1},
\end{gathered}
}[EqMoments]
respectively.  Here $\mathcal{V}_N^2$ is the Stiefeld manifold for the group of $N\times N$ unitary matrices $U(N)$ and its volume is given by
\eqn{\text{Vol}\left(\mathcal{V}_N^2\right)=\int_{U\in\mathcal{V}_N^2}U^\dagger dU=\frac{2^N\pi^{N(N+1)/2}}{\prod_{1\leq i\leq N}\Gamma(i)}.}

First, we introduce the permutation matrices
\eqn{(\Pi_{ab})_{ij}=\delta_{ij}-\delta_{ia}\delta_{aj}-\delta_{ib}\delta_{bj}+\delta_{ia}\delta_{bj}+\delta_{ib}\delta_{aj}.}[EqPerm]
It is easy to see that $\Pi_{ab}M$ permutes the $a$-th and $b$-th rows of $M$ while $M\Pi_{ab}$ permutes the $a$-th and $b$-th columns of $M$.  Since $\Pi_{ab}^\dagger=\Pi_{ab}$, then $U\to\Pi_{ab}U\Pi_{cd}$ is unitary and the Haar measure does not change, $U^\dagger dU\to U^\dagger dU$.

Therefore, with an appropriate change of integration variables using the permutation matrices \eqref{EqPerm}, we have
\eqna{
\left\langle(x_{ij})^{s-1}\right\rangle&=\frac{1}{\text{Vol}\left(\mathcal{V}_N^2\right)}\int_{U\in\mathcal{V}_N^2}U^\dagger dU\,|U_{ij}|^{2(s-1)}=\frac{1}{\text{Vol}\left(\mathcal{V}_N^2\right)}\int_{U\in\mathcal{V}_N^2}U^\dagger dU\,|(\Pi_{ia}U\Pi_{jb})_{ij}|^{2(s-1)}\\
&=\frac{1}{\text{Vol}\left(\mathcal{V}_N^2\right)}\int_{U\in\mathcal{V}_N^2}U^\dagger dU\,|U_{ab}|^{2(s-1)}=\left\langle(x_{ab})^{s-1}\right\rangle,
}
as well as
\eqna{
\left\langle(x_j^M)^{s-1}\right\rangle&=\frac{1}{\text{Vol}\left(\mathcal{V}_N^2\right)}\int_{U\in\mathcal{V}_N^2}U^\dagger dU\,|4\text{Im}(U_{i_0j}U_{i_0j}U_{i_0j_0}^*U_{i_0j_0}^*)|^{2(s-1)}\\
&=\frac{1}{\text{Vol}\left(\mathcal{V}_N^2\right)}\int_{U\in\mathcal{V}_N^2}U^\dagger dU\,|4\text{Im}[(U\Pi_{jb})_{i_0j}(U\Pi_{jb})_{i_0j}(U\Pi_{jb})_{i_0j_0}^*(U\Pi_{jb})_{i_0j_0}^*]|^{2(s-1)}\\
&=\frac{1}{\text{Vol}\left(\mathcal{V}_N^2\right)}\int_{U\in\mathcal{V}_N^2}U^\dagger dU\,|4\text{Im}(U_{i_0b}U_{i_0b}U_{i_0j_0}^*U_{i_0j_0}^*)|^{2(s-1)}=\left\langle(x_b^M)^{s-1}\right\rangle,
}
and finally
\eqna{
\left\langle(x_{ij}^D)^{s-1}\right\rangle&=\frac{1}{\text{Vol}\left(\mathcal{V}_N^2\right)}\int_{U\in\mathcal{V}_N^2}U^\dagger dU\,|6\sqrt{3}\text{Im}(U_{i_0j_0}U_{ij}U_{i_0j}^*U_{ij_0}^*)|^{2(s-1)}\\
&=\frac{1}{\text{Vol}\left(\mathcal{V}_N^2\right)}\int_{U\in\mathcal{V}_N^2}U^\dagger dU\,|6\sqrt{3}\text{Im}[(\Pi_{ia}U\Pi_{jb})_{i_0j_0}(\Pi_{ia}U\Pi_{jb})_{ij}(\Pi_{ia}U\Pi_{jb})_{i_0j}^*(\Pi_{ia}U\Pi_{jb})_{ij_0}^*]|^{2(s-1)}\\
&=\frac{1}{\text{Vol}\left(\mathcal{V}_N^2\right)}\int_{U\in\mathcal{V}_N^2}U^\dagger dU\,|6\sqrt{3}\text{Im}(U_{i_0j_0}U_{ab}U_{i_0b}^*U_{aj_0}^*)|^{2(s-1)}=\left\langle(x_{ab}^D)^{s-1}\right\rangle.
}
Again, in each of these equations, we simply implemented a change of integration variables, changing $U\to\Pi_{ab}U\Pi_{cd}$ with the appropriate $a$, $b$, $c$, and $d$.  Moreover, we relied on the left- and right-invariance of the Haar measure.  Also, we note that since $i,j\neq i_0,j_0$, the indices $i_0$ and $j_0$ did not change under the permutations.

We now conclude that under the Haar measure, the moments \eqref{EqMoments} of the rephasing invariants of the same type are all equal.  Since the PDF is completely determined by its moments, this demonstration implies that all the rephasing invariants of a particular type have the same PDF.  Therefore, there are only three PDFs to determine: one for the quadratic rephasing invariants $x$, one for the quartic Majorana rephasing invariants $x^M$, and one for the quartic Dirac rephasing invariants $x^D$.


\section{Mellin Transform}\label{SecMellin}

This section reviews the Mellin transform.  We first discuss in all generality how to compute PDFs from their moments with the help of the Mellin transform.  We then focus on moments of the particular type that occur for our rephasing invariants and express the relevant PDFs in terms of Meijer $G$-functions.


\subsection{Mellin Transform Method}

The Mellin transform of a function $f(x_i)$ is defined as
\eqn{\{\mathcal{M}f\}(s_1,\ldots,s_n)\equiv\int_0^\infty\left[\prod_{k=1}^ndx_k\,x_k^{s_k-1}\right]f(x_1,\ldots,x_n)=g(s_1,\ldots,s_n),}[EqMellin]
where the Mellin transform $g(s_1,\ldots,s_n)$ is a function of the variables $s_i$, the conjugate variables associated to the $x_i$.  The inverse Mellin transform is given by
\eqn{\{\mathcal{M}^{-1}g\}(x_1,\ldots,x_n)\equiv\int_{\gamma-i\infty}^{\gamma+i\infty}\left[\prod_{k=1}^n\frac{ds_k}{2\pi i}\,x_k^{-s_k}\right]g(s_1,\ldots,s_n)=f(x_1,\ldots,x_n),}[EqInvMellin]
for an appropriate choice of $\gamma$.

The Mellin transform \eqref{EqMellin} is a powerful tool to determine a PDF from the knowledge of its moments.  Indeed for an unknown PDF $f(x_1,\ldots,x_n)$ of $n$ random variables $x_i$ with support on the positive axes, by definition the Mellin transform $g(s_1,\ldots,s_n)$ corresponds to its moments.  Hence, it is possible to obtain the unknown PDF by operating an inverse Mellin transform \eqref{EqInvMellin} on the moments.

More precisely, the moments, which are given by
\eqn{\left\langle x_1^{s_1-1}\cdots x_n^{s_n-1}\right\rangle=\int_0^\infty\left[\prod_{k=1}^ndx_k\,x_k^{s_k-1}\right]f(x_1,\ldots,x_n)=\{\mathcal{M}f\}(s_1,\ldots,s_n),}
are simply the Mellin transform \eqref{EqMellin}.  Therefore, the inverse Mellin transform \eqref{EqInvMellin} of the moments
\eqn{\left\{\mathcal{M}^{-1}\left\langle x_1^{s_1-1}\cdots x_n^{s_n-1}\right\rangle\right\}(x_1,\ldots,x_n)=f(x_1,\ldots,x_n),}
leads directly to the PDF of interest $f(x_1,\ldots,x_n)$.


\subsection{Meijer $G$-functions and Generalized Harmonic Numbers}

In the computation of the moments \eqref{EqMoments} from the explicit Haar measure \eqref{EqHaar}, we come across $\xi_k=[1+(-1)^{2k}]/2$ and the moments
\eqn{\left\langle x^n\right\rangle=\prod_{k=1}^m\frac{\Pochhammer{\alpha_k}{n}}{\Pochhammer{\alpha_k+\beta_k}{n}},}[EqMomentsPoch]
where the $\alpha_k$ and $\beta_k$ are real and positive (see \cite{Fortin:2018etr} for more detail).  We thus investigate the PDF associated to the moments \eqref{EqMomentsPoch} before proceeding with the explicit moments for the rescaled rephasing invariants \eqref{EqRIscaled}.

From the discussion above, the PDF $f(x)$ for the moments \eqref{EqMomentsPoch} is simply the inverse Mellin transform \eqref{EqInvMellin}, which gives
\eqna{
f(x)&=\left\{\mathcal{M}^{-1}\left\langle x^{s-1}\right\rangle\right\}(x)=\left[\prod_{k=1}^m\frac{\Gamma(\alpha_k+\beta_k)}{\Gamma(\alpha_k)}\right]\frac{1}{2\pi i}\int_{\gamma-i\infty}^{\gamma+i\infty}ds\,x^{-s}\prod_{k=1}^m\frac{\Gamma(\alpha_k-1+s)}{\Gamma(\alpha_k+\beta_k-1+s)}\\
&=\left[\prod_{k=1}^m\frac{\Gamma(\alpha_k+\beta_k)}{\Gamma(\alpha_k)}\right]G_{m,m}^{m,0}\left(\left.\arraycolsep=1pt\def\arraystretch{1}\begin{array}{cccccc}\alpha_1+\beta_1-1,&\ldots,&\alpha_m+\beta_m-1\\\alpha_1-1,&\ldots,&\alpha_m-1\end{array}\right|x\right),
}[EqPDFPoch]
where the last equality necessitates $\sum_{1\leq i\leq m}\beta_i<-1$ for convergence.  This result is expressed in terms of the Meijer $G$-function
\eqn{G_{p,q}^{m,n}\left(\left.\arraycolsep=1pt\def\arraystretch{1}\begin{array}{cccccc}a_1,&\ldots,&a_n,&a_{n+1},&\ldots,&a_p\\b_1,&\cdots,&b_m,&b_{m+1},&\ldots,&b_q\end{array}\right|z\right)=\frac{1}{2\pi i}\int_\mathcal{L}ds\,z^{-s}\frac{\prod_{k=1}^m\Gamma(s+b_k)\prod_{k=1}^n\Gamma(1-a_k -s)}{\prod_{k=n+1}^p\Gamma(s+a_k)\prod_{k=m+1}^q\Gamma(1-b_k-s)},}[EqMeijer]
where $\mathcal{L}$ is the proper contour.

In the analysis of the behavior of the Meijer $G$-function \eqref{EqMeijer} around the origin, we encounter the generalized harmonic numbers $H_{n,m}$, which are defined as
\eqn{H_{n,m}=\sum_{k=1}^n\frac{1}{k^m},\qquad H_n\equiv H_{n,1}.}[EqGHN]

Therefore, the PDF for the moments \eqref{EqMomentsPoch} is simply given by \eqref{EqPDFPoch} which is written in terms of the Meijer $G$-function \eqref{EqMeijer}, and its behavior around the origin lead to the generalized harmonic numbers \eqref{EqGHN}.


\section{Rephasing Invariant PDFs for Arbitrary Neutrino Number}\label{SecPDF}

In this section, we finally determine the three different PDFs for the rephasing invariants using the results of the previous sections.  For each case, we first find the simplest rephasing invariant with the parametrization \eqref{EqU} and use the Haar measure \eqref{EqHaar} to determine the moments (see \cite{Fortin:2018etr}).  Then we find the associated PDF with the help of the inverse Mellin transform \eqref{EqInvMellin}.  For the quartic rephasing invariants, the results are expressed in terms of Meijer $G$-functions \eqref{EqMeijer}.


\subsection{Quadratic Invariants}

The simplest quadratic rephasing invariant in the parametrization \eqref{EqU} appears when we set $i=N$ and $j=1$.  In that case, we have
\eqn{x=|U_{N1}|^2=\sin^2(\theta_{1N}),\qquad0\leq x\leq1.}
With the Haar measure \eqref{EqHaar}, the moments are easily computed and are given by
\eqn{\left\langle x^{s-1}\right\rangle=\frac{\Gamma(N)\Gamma(s)}{\Gamma(N-1+s)}.}
Hence, from \eqref{EqPDFPoch} the PDF is
\eqn{\mathcal{P}(x)dx=\left\{\mathcal{M}^{-1}\left\langle x^{s-1}\right\rangle\right\}(x)dx=(N-1)(1-x)^{N-2}dx,}[EqPDFQuad]
for all quadratic rephasing invariants.


\subsection{Quartic Majorana Invariants}

In the parametrization \eqref{EqU}, the simplest quartic Majorana invariant is obtained by setting $i_0=N$, $j_0=1$, and $j=2$.  From \eqref{EqRI}, the rephasing invariant takes the form
\eqna{
y^M&=\text{Im}(U_{N2}U_{N2}U_{N1}^*U_{N1}^*)\\
&=\left\{\begin{array}{cc}\cos^2(\theta_{12})\sin^2(\theta_{12})\sin(-2\varphi_1+2\varphi_2)&N=2\\\cos^2(\theta_{1N})\sin^2(\theta_{1N})\sin^2(\theta_{2N})\sin(2\phi_{2N}-2\varphi_1+2\varphi_2)&N>2\end{array}\right.,
}
with the rephasing invariant defined in the interval
\eqn{-\frac{1}{4}\leq y^M\leq\frac{1}{4}.}
Clearly, the odd moments under the Haar measure vanish, justifying the switch to the rescaled quartic Majorana invariant \eqref{EqRIscaled}.

A direct computation with the explicit form of the Haar measure leads to the moments
\eqn{\left\langle(x^M)^{s-1}\right\rangle=\frac{2^{\frac{5}{2}-2N}\xi_{s-1}\Gamma(N)\Gamma\left(s-\frac{1}{2}\right)^3\Gamma(s)}{\Gamma\left(s+\frac{N-4}{4}\right)\Gamma\left(s+\frac{N-3}{4}\right)\Gamma\left(s+\frac{N-2}{4}\right)\Gamma\left(s+\frac{N-1}{4}\right)},\qquad\forall\,N\geq2.}
Thus, using the inverse Mellin transform and \eqref{EqPDFPoch}, the PDF for the rescaled quartic Majorana rephasing invariant \eqref{EqRIscaled} can be expressed as
\eqna{
\mathcal{P}_M(x^M)dx^M&=\left\{\mathcal{M}^{-1}\left\langle(x^M)^{s-1}\right\rangle\right\}(x^M)dx^M\\
&=2^{\frac{5}{2}-2N}\Gamma(N)G_{4,4}^{4,0}\left(\left.\arraycolsep=1pt\def\arraystretch{1}\begin{array}{ccccccc}\frac{N-4}{4},&\frac{N-3}{4},&\frac{N-2}{4},&\frac{N-1}{4}\\-\frac{1}{2},&-\frac{1}{2},&-\frac{1}{2},&0\end{array}\right|x^M\right)dx^M,
}
or, in terms of the quartic Majorana rephasing invariant $y^M$,
\eqn{\mathcal{P}_M(y^M)dy^M=2^{\frac{13}{2}-2N}\Gamma(N)|y^M|G_{4,4}^{4,0}\left(\left.\arraycolsep=1pt\def\arraystretch{1}\begin{array}{ccccccc}\frac{N-4}{4},&\frac{N-3}{4},&\frac{N-2}{4},&\frac{N-1}{4}\\-\frac{1}{2},&-\frac{1}{2},&-\frac{1}{2},&0\end{array}\right|16|y^M|^2\right)dy^M,}[EqPDFQuarM]
for all quartic Majorana rephasing invariants.


\subsection{Quartic Dirac Invariants}

Finally, the simplest quartic Dirac invariant in the parametrization \eqref{EqU} originates from setting $i_0=N-1$, $j_0=2$, $i=N$, and $j=1$.  With this choice, the associated rephasing invariant \eqref{EqRI} is expressed as
\eqna{
y^D&=\text{Im}(U_{N-1,2}U_{N1}U_{N-1,1}^*U_{N2}^*)\\
&=\left\{\begin{array}{cc}\substack{\cos^2(\theta_{13})\sin(\theta_{13})\cos(\theta_{12})\sin(\theta_{12})\\\times\cos(\theta_{23})\sin(\theta_{23})\sin(\phi_{23})}&N=3\\[10pt]\substack{\cos^2(\theta_{1N})\sin(\theta_{1N})\cos(\theta_{1,N-1})\sin(\theta_{1,N-1})\cos(\theta_{2N})\\\times\sin(\theta_{2N})\sin(\theta_{2,N-1})\sin(\phi_{2,N-1}-\phi_{2N})}&N>3\end{array}\right.,
}
with the rephasing invariant defined in the interval
\eqn{-\frac{1}{6\sqrt{3}}\leq y^D\leq\frac{1}{6\sqrt{3}}.}
Once again, we can directly see that under the Haar measure, the odd moments vanish.  This observation justifies using the rescaled quartic Dirac invariant \eqref{EqRIscaled}.

Following \cite{Fortin:2018etr} with the help of the Haar measure \eqref{EqHaar}, the moments are easily computed to be
\eqn{\left\langle(x^D)^{s-1}\right\rangle=\frac{2^{3-N}3^{\frac{1}{2}-N}\pi\xi_{s-1}(N-2)\Gamma(N)\Gamma\left(s-\frac{1}{2}\right)\Gamma(s)^3\Gamma(s+N-3)}{\Gamma\left(s+\frac{N-3}{2}\right)\Gamma\left(s+\frac{N-2}{2}\right)\Gamma\left(s+\frac{N-3}{3}\right)\Gamma\left(s+\frac{N-2}{3}\right)\Gamma\left(s+\frac{N-1}{3}\right)},\qquad\forall\,N\geq3.}
From \eqref{EqPDFPoch}, the rescaled quartic Dirac rephasing invariant \eqref{EqRIscaled} has for PDF
\eqna{
\mathcal{P}_D(x^D)dx^D&=\left\{\mathcal{M}^{-1}\left\langle(x^D)^{s-1}\right\rangle\right\}(x^D)dx^D\\
&=2^{3-N}3^{\frac{1}{2}-N}\pi(N-2)\Gamma(N)G_{5,5}^{5,0}\left(\left.\arraycolsep=1pt\def\arraystretch{1}\begin{array}{cccccccc}\frac{N-3}{2},&\frac{N-2}{2},&\frac{N-3}{3},&\frac{N-2}{3},&\frac{N-1}{3}\\-\frac{1}{2},&0,&0,&0,&N-3\end{array}\right|x^D\right)dx^D,
}
which translates into
\eqn{\mathcal{P}_D(y^D)dy^D=2^{5-N}3^{\frac{7}{2}-N}\pi(N-2)\Gamma(N)|y^D|G_{5,5}^{5,0}\left(\left.\arraycolsep=1pt\def\arraystretch{1}\begin{array}{cccccccc}\frac{N-3}{2},&\frac{N-2}{2},&\frac{N-3}{3},&\frac{N-2}{3},&\frac{N-1}{3}\\-\frac{1}{2},&0,&0,&0,&N-3\end{array}\right|108|y^D|^2\right)dy^D,}[EqPDFQuarD]
for all quartic Dirac rephasing invariants $y^D$.


\section{Discussion}\label{SecDisc}

This section compares the analytic PDFs obtained above with numerical results, investigates the behavior of the PDFs around the origin (vanishing rephasing invariants), and discusses the physical implications of the PDFs (considering $|y|$ instead of $y$ for the quartic rephasing invariants due to their PDF invariance under $y\to-y$).


\subsection{Analysis of the PDFs}

The three PDFs in function of the neutrino number $N$ for the quadratic, quartic Majorana, and quartic Dirac rephasing invariants are given in \eqref{EqPDFQuad}, \eqref{EqPDFQuarM}, and \eqref{EqPDFQuarD}, respectively.  We reproduce the results here for convenience:
\eqn{
\begin{gathered}
\mathcal{P}(x)dx=\left\{\mathcal{M}^{-1}\left\langle x^{s-1}\right\rangle\right\}(x)dx=(N-1)(1-x)^{N-2}dx,\\
\mathcal{P}_M(y^M)dy^M=2^{\frac{13}{2}-2N}\Gamma(N)|y^M|G_{4,4}^{4,0}\left(\left.\arraycolsep=1pt\def\arraystretch{1}\begin{array}{ccccccc}\frac{N-4}{4},&\frac{N-3}{4},&\frac{N-2}{4},&\frac{N-1}{4}\\-\frac{1}{2},&-\frac{1}{2},&-\frac{1}{2},&0\end{array}\right|16|y^M|^2\right)dy^M,\\
\mathcal{P}_D(y^D)dy^D=2^{5-N}3^{\frac{7}{2}-N}\pi(N-2)\Gamma(N)|y^D|G_{5,5}^{5,0}\left(\left.\arraycolsep=1pt\def\arraystretch{1}\begin{array}{cccccccc}\frac{N-3}{2},&\frac{N-2}{2},&\frac{N-3}{3},&\frac{N-2}{3},&\frac{N-1}{3}\\-\frac{1}{2},&0,&0,&0,&N-3\end{array}\right|108|y^D|^2\right)dy^D.
\end{gathered}
}[EqPDF]
We can now compare the analytic results \eqref{EqPDF} with numerical results and investigate the behavior of the PDFs \eqref{EqPDF} around the origin.  For the numerical results with a given $N$, we simply generate a large sample of random $N\times N$ unitary matrices and determine their rephasing invariants.  In each case (quadratic, quartic Majorana, and quartic Dirac), we did verify numerically that all rephasing invariants of the same type have the same PDF.

\begin{figure}[!t]
\centering
\resizebox{15cm}{!}{
\includegraphics{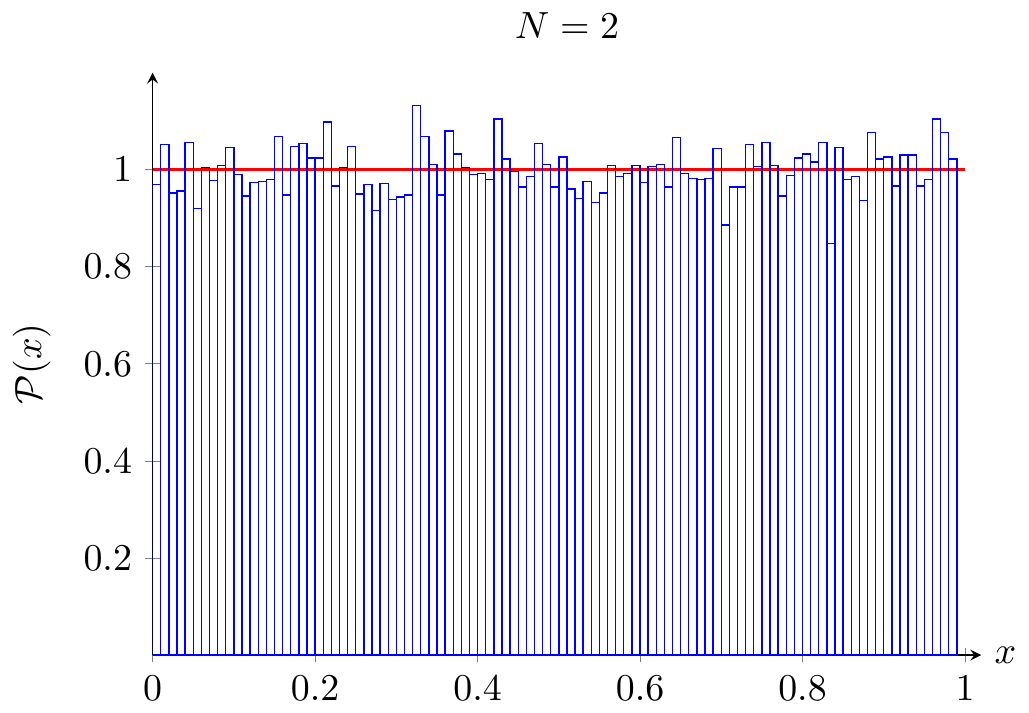}
\hspace{2cm}
\includegraphics{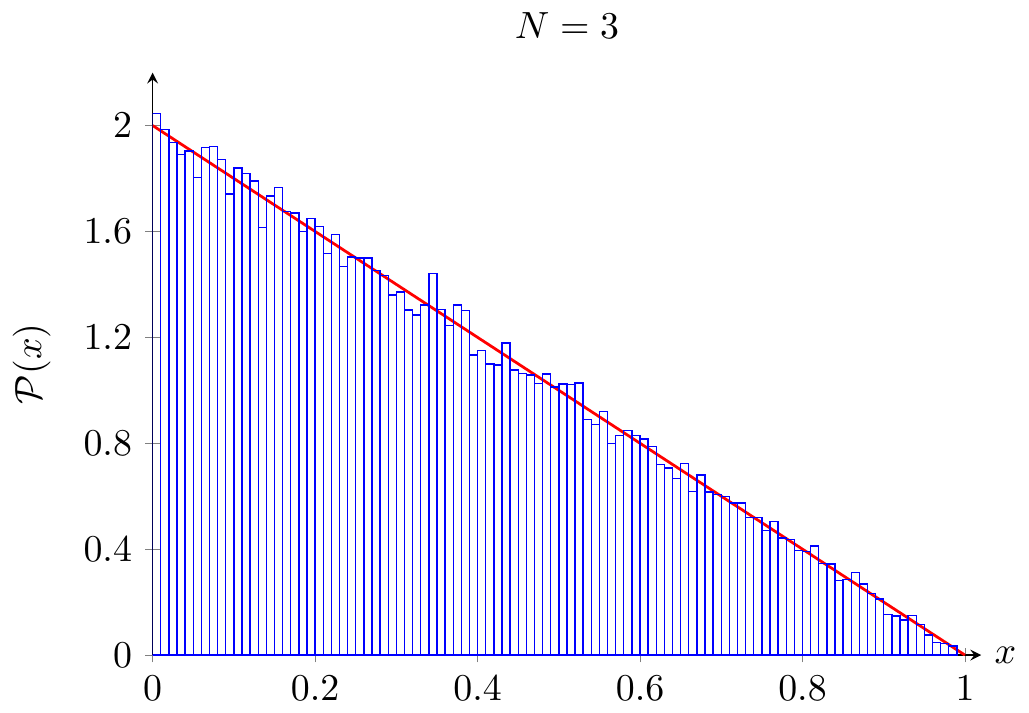}
}\\
\vspace{1cm}
\resizebox{15cm}{!}{
\includegraphics{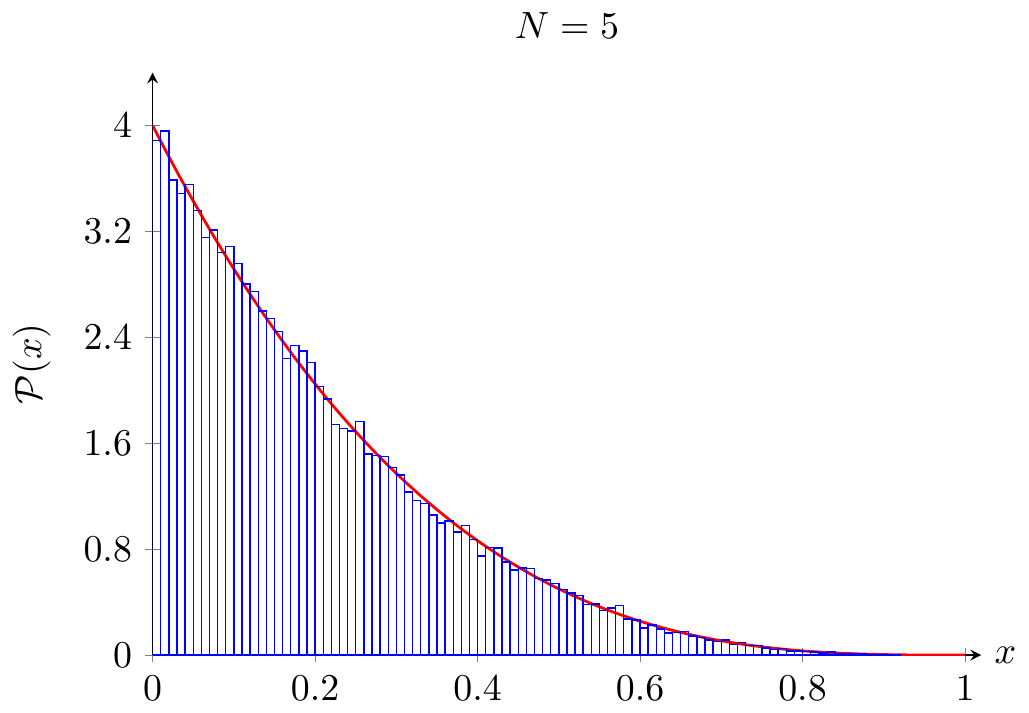}
\hspace{2cm}
\includegraphics{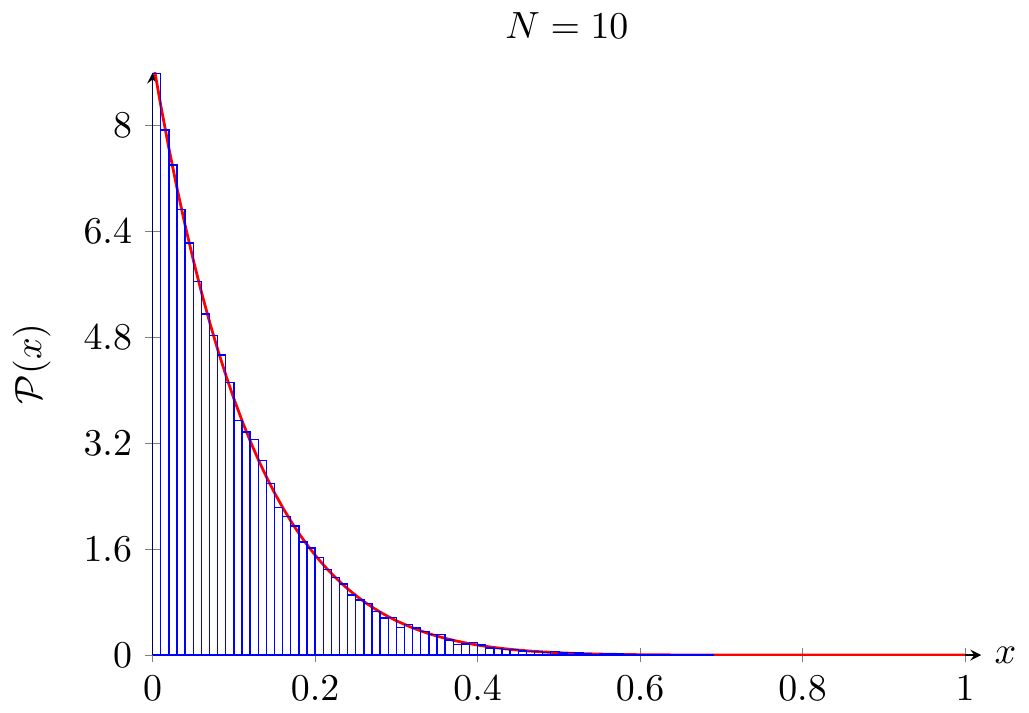}
}
\caption{Quadratic rephasing invariant PDFs for different values of $N$.  The red curves correspond to the analytic results while the histograms correspond to the numerical results with a sample of $5\times10^4$ unitary matrices.}
\label{FigPDFQuad}
\end{figure}

We first begin with the quadratic rephasing invariant.  Their PDFs \eqref{EqPDFQuad} for different $N$ are shown in Figure \ref{FigPDFQuad}
and their behavior around $x=0$ is given by
\eqn{\mathcal{P}(x)\sim(N-1)[1-(N-2)x].}
Clearly, the quadratic rephasing invariant PDFs peak around $x=0$ as the number of neutrinos $N$ increases.  This feature is common to all types of rephasing invariants.

\begin{figure}[!t]
\centering
\resizebox{15cm}{!}{
\includegraphics{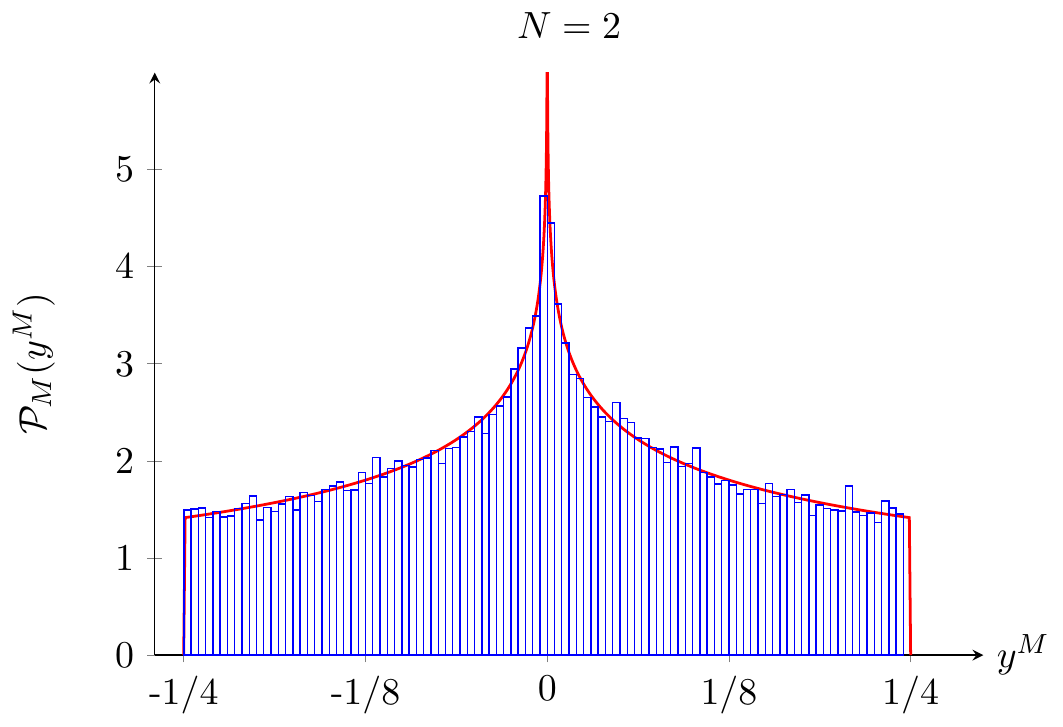}
\hspace{2cm}
\includegraphics{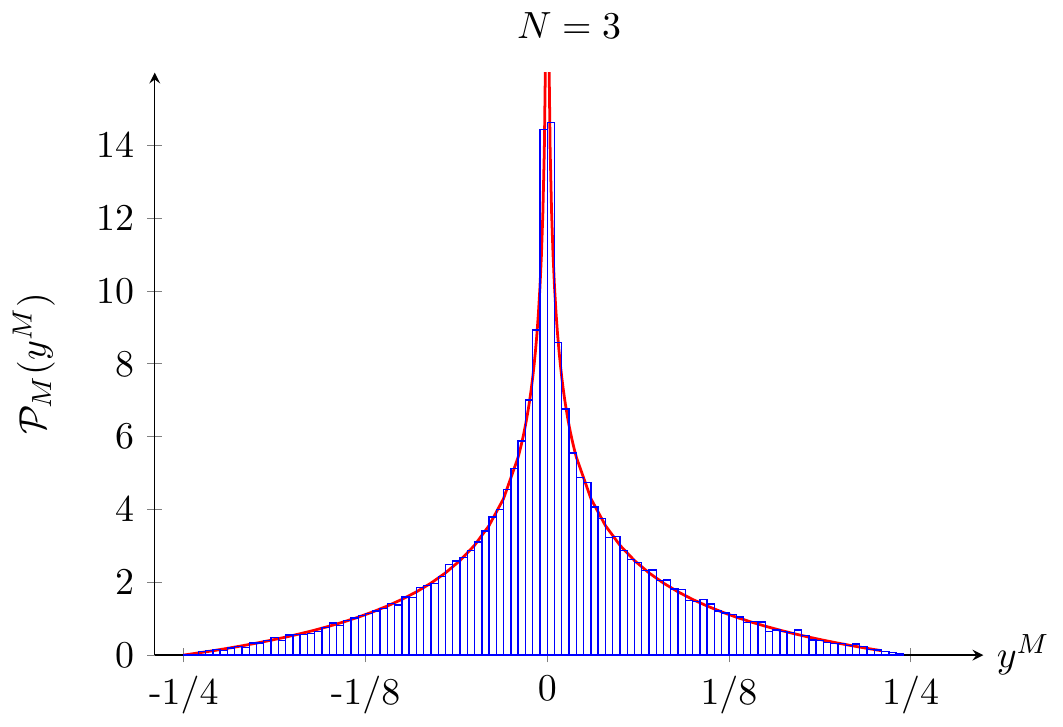}
}\\
\vspace{1cm}
\resizebox{15cm}{!}{
\includegraphics{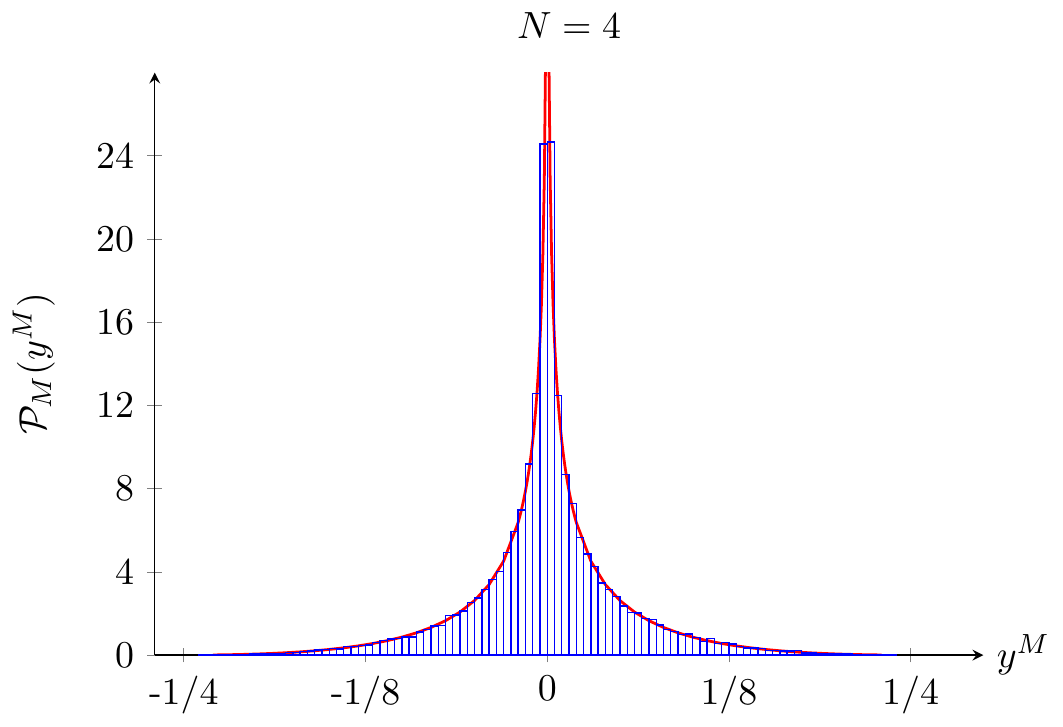}
\hspace{2cm}
\includegraphics{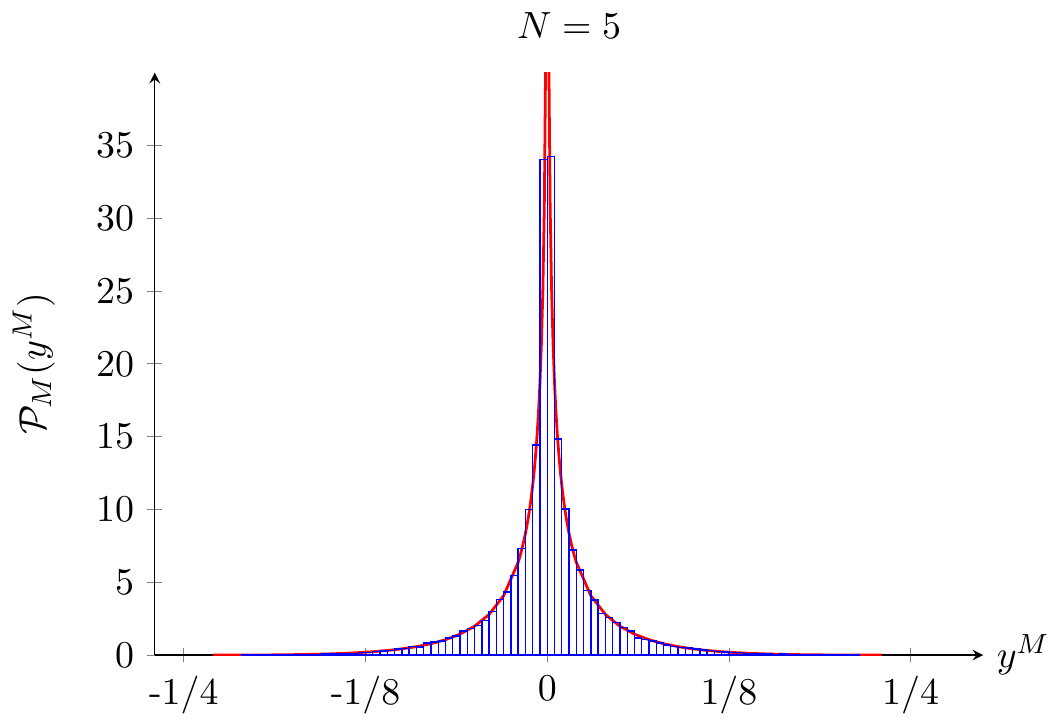}
}
\caption{Quartic Majorana rephasing invariant PDFs for different values of $N$.  The red curves correspond to the analytic results while the histograms correspond to the numerical results with a sample of $5\times10^4$ unitary matrices.}
\label{FigPDFQuarM}
\end{figure}

The quartic Majorana rephasing invariant PDFs \eqref{EqPDFQuarM} for different $N$ are shown in Figure \ref{FigPDFQuarM}.
Their behavior around $y^M=0$ can be written as
\eqna{
\mathcal{P}_M(y^M)&\sim\frac{(N-2)(N-1)}{8\pi}\left[\ln^2\left(16|y^M|^2\right)+4(2H_{N-3}-3\ln2)\ln\left(16|y^M|^2\right)\right.\\
&\phantom{\sim}\qquad\left.-12(4H_{N-3}-3\ln2)\ln2+16H_{N-3}^2+16H_{N-3,2}-\frac{5\pi^2}{3}\right],
}
where we used \eqref{EqGHN}.  We note that the case $N=2$ must be evaluated with the help of the limit $N\to2$.  Moreover, contrary to the two other PDFs, the PDF for the quartic Majorana rephasing invariants blows up at the origin.

\begin{figure}[!t]
\centering
\resizebox{15cm}{!}{
\includegraphics{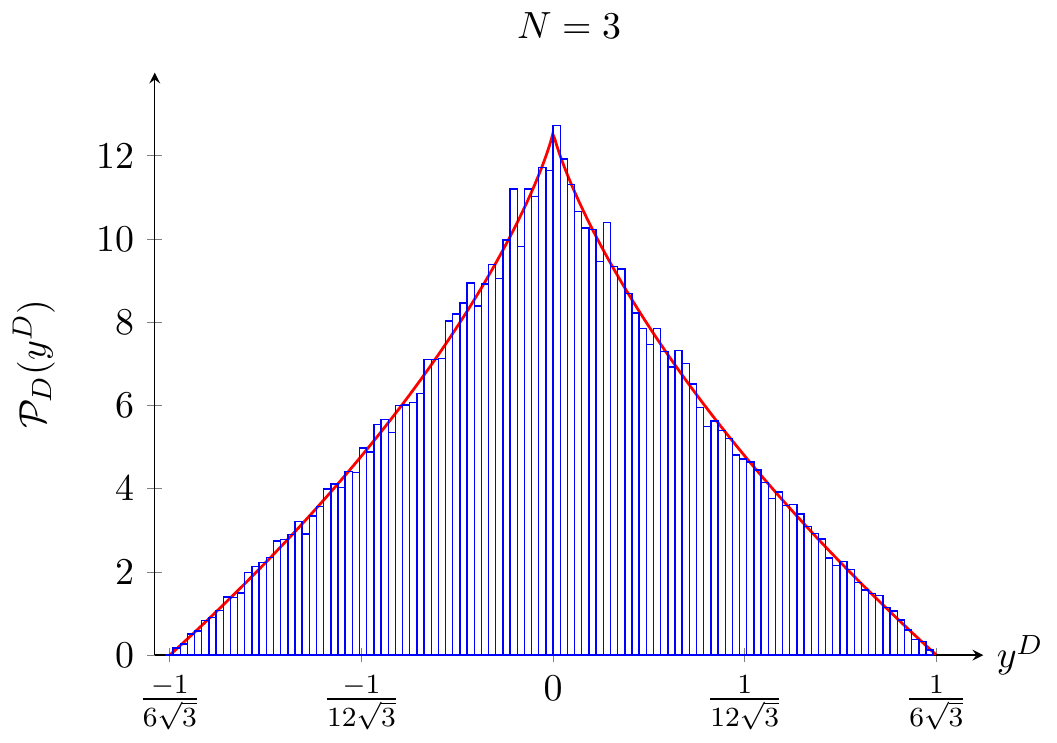}
\hspace{2cm}
\includegraphics{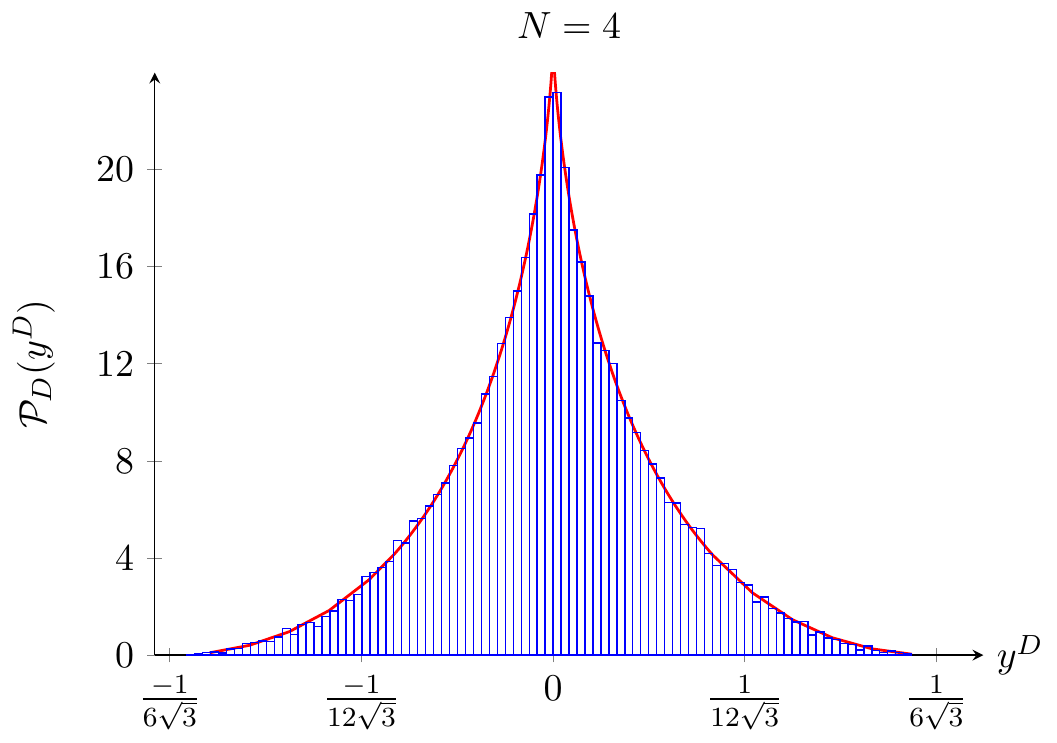}
}\\
\vspace{1cm}
\resizebox{15cm}{!}{
\includegraphics{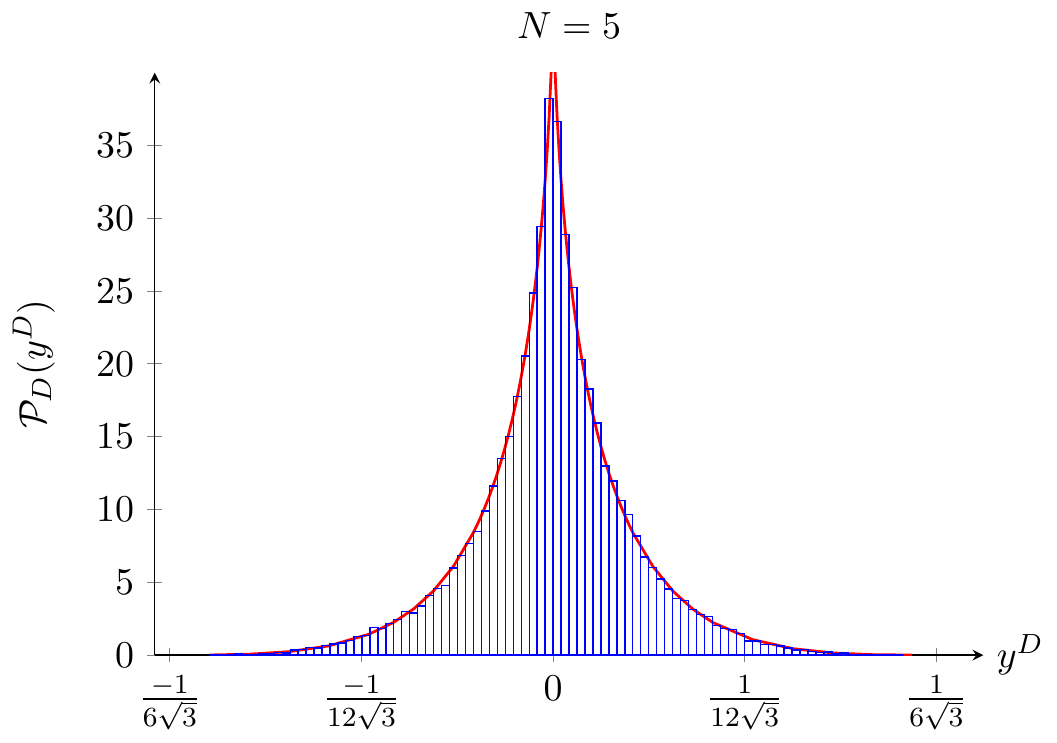}
\hspace{2cm}
\includegraphics{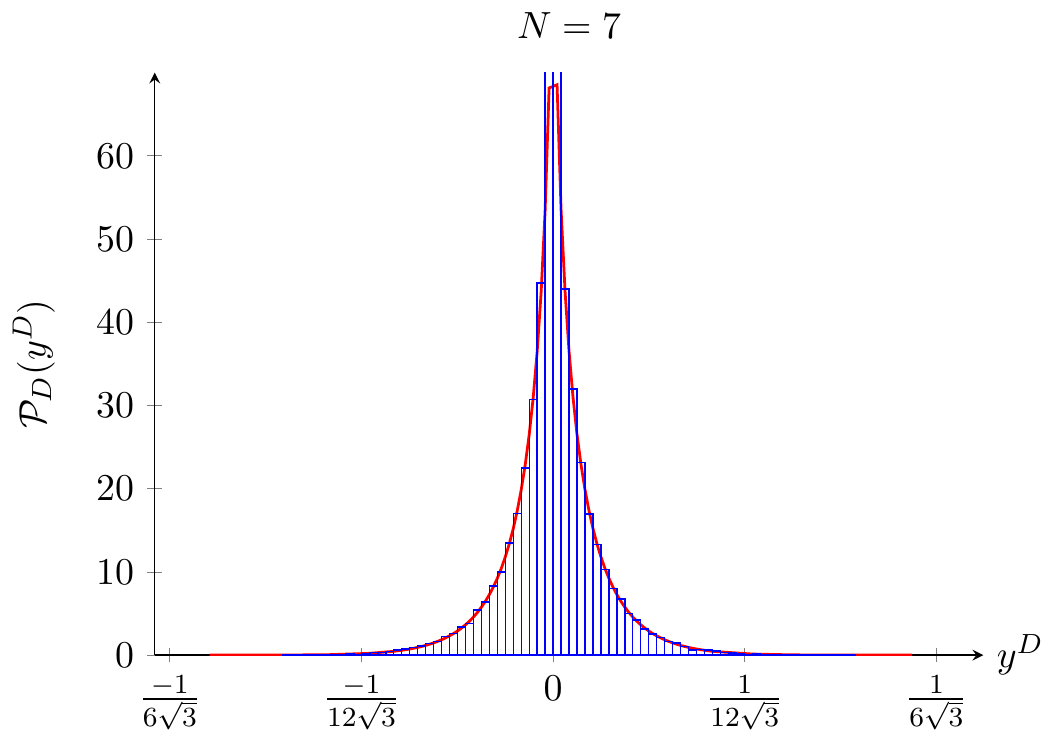}
}
\caption{Quartic Dirac rephasing invariant PDFs for different values of $N$.  The red curves correspond to the analytic results while the histograms correspond to the numerical results with a sample of $5\times10^4$ unitary matrices.}
\label{FigPDFQuarD}
\end{figure}

Finally, for different choices of $N$, the quartic Dirac rephasing invariant PDFs \eqref{EqPDFQuarD} are illustrated in Figure \ref{FigPDFQuarD}.
Using \eqref{EqGHN} again, around the origin $y^D=0$ the PDFs behave as
\eqn{\mathcal{P}_D(y^D)\sim4\pi+24|y^D|\left[\ln^2\left(108|y^D|^2\right)-3\ln3-2\right],}
for $N=3$ and
\eqna{
\mathcal{P}_D(y^D)&\sim\frac{2\pi(N-2)^2(N-1)}{2N-5}-(N-3)(N-2)^2(N-1)|y^D|\\
&\phantom{\sim}\qquad\times\left[\ln^2\left(108|y^D|^2\right)+2(4H_{N-4}-2-3\ln3)\ln\left(108|y^D|^2\right)\right.\\
&\phantom{\sim}\qquad\left.-3(8H_{N-4}-4-3\ln3)\ln3+16H_{N-4}^2-16H_{N-4}+12H_{N-4,2}+8-\pi^2\right],
}
for $N>3$.  The case $N=3$ must be considered separately since the limit $N\to3$ does not commute with the limit $y^D\to0$.

We note that the analytic results \eqref{EqPDF} are in perfect agreement with the numerical results, validating our approach based on the Mellin transform.  Moreover, although they are not expressed in the same way, we have checked that the explicit PDFs \eqref{EqPDF} match the ones found in \cite{Fortin:2018etr} for $N=2$ and $N=3$.\footnote{The equality of the PDFs implies identities between the Meijer $G$-functions obtained here and the expressions in terms of hypergeometric functions and Meijer $G$-functions computed in \cite{Fortin:2018etr}.}


\subsection{Analysis of the Average Values}

By analyzing the PDFs and the average values of $|y^M|$ and $|y^D|$, it was argued in \cite{Fortin:2018etr} that the $3\times3$ PMNS matrix was likely to have been drawn randomly from a probability experiment distributed following the Haar measure.  Moreover, it was found that the $N=3$ average value $\left\langle|y^D|\right\rangle=\pi/105\approx0.030$ was in very good agreement with the experimental value $|y_\text{exp}^D|=0.032\pm0.005$.  It is of interest here to investigate the average values for arbitrary neutrino number $N$, which could be relevant for physics beyond the Standard Model with sterile neutrinos.

\begin{figure}[!t]
\centering
\resizebox{15cm}{!}{
\includegraphics{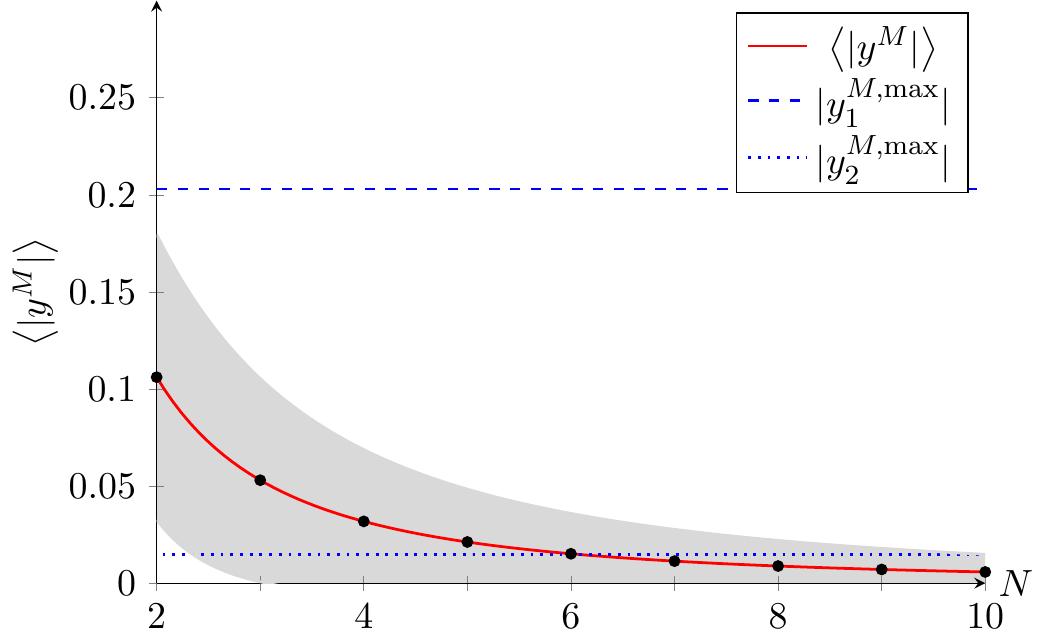}
\hspace{2cm}
\includegraphics{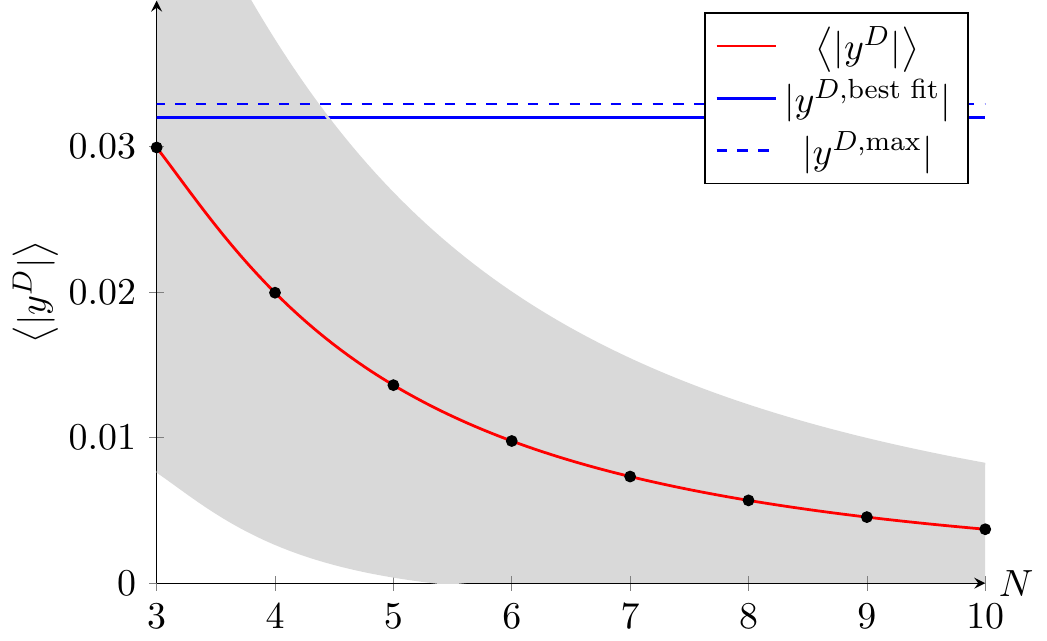}
}
\caption{Quartic rephasing invariant average values (solid red lines and black dots, with shaded regions representing one standard deviation) in function of the neutrino number for Majorana (left panel) and Dirac (right panel) rephasing invariants.  In both panels, the dashed and dotted blue lines represent the maximum allowed values (for $N=3$) calculated from the experimental values for the mixing angles and phases.  In the right panel, the solid blue line represents the best-fit observed value (for $N=3$) calculated from the experimental values for the mixing angles and phases.}
\label{FigAvg}
\end{figure}

Using \eqref{EqPDF} or the associated moments, the average values of the absolute values of the rephasing invariants and the average values of the rephasing invariants square are given by
\eqn{
\begin{gathered}
\left\langle x\right\rangle=\frac{1}{N},\qquad\left\langle x^2\right\rangle=\frac{2}{N(N+1)},\\
\left\langle|y^M|\right\rangle=\frac{2}{\pi N(N+1)},\qquad\left\langle|y^M|^2\right\rangle=\frac{2}{N(N+1)(N+2)(N+3)},\\
\left\langle|y^D|\right\rangle=\frac{\pi(N-2)}{(2N-3)(2N-1)(2N+1)},\qquad\left\langle|y^D|^2\right\rangle=\frac{N-2}{2(N-1)N^2(N+1)(N+2)}.
\end{gathered}
}[EqAvg]
A comparison of the averages \eqref{EqAvg} as a function of the neutrino number $N$ and the experimental values is provided in Figure \ref{FigAvg}.
Here the solid red lines are the average values \eqref{EqAvg} while the shaded regions represent one standard deviation away from the average values, also computed from \eqref{EqAvg}.  Moreover, the dashed and dotted blue lines correspond to the maximum allowed values calculated from the experimental values for the mixing angles and phases.  For the two quartic Majorana rephasing invariants, the values correspond to $y_1^M$ (dashed) and $y_2^M$ (dotted) respectively while for the Dirac rephasing invariant, the solid blue line corresponds to the best-fit observed value.  We see that under the probabilistic approach used here with the Haar measure, the case $N=3$ is the best case scenario to match with Nature when considering the quartic Dirac rephasing invariants.  For the quartic Majorana invariants, our statistical approach also points toward $N=3$ when considering the maximum allowed values for both invariants ($N=2$ would be better, since it allows for the very large $|y_1^M|$, but that case is excluded).  We thus conclude that in our framework, there would not be any extra sterile neutrino (apart \textit{e.g.} from the three heavy neutrinos responsible for the type I seesaw mechanism).

Before concluding, it is of interest to point out that the largest rephasing invariants obtained from the Haar measure originate from the smallest neutrino number.  Hence CP violation is larger for smaller $N$.  This matches with our observation that all three PDFs peak around the origin as the number of neutrinos $N$ increases, leading to vanishing moments as $N\to\infty$.  In fact, it is now easy to perform a large $N$ analysis.  For example, from the average values \eqref{EqAvg}, we see that
\eqn{\left\langle x\right\rangle=\frac{1}{N},\qquad\left\langle|y^M|\right\rangle\approx\frac{2}{\pi N^2}\left[1-\frac{1}{N}+\cdots\right],\qquad\left\langle|y^D|\right\rangle\approx\frac{\pi}{8N^2}\left[1-\frac{1}{2N}+\cdots\right].}[EqAvgLargeN]
Therefore, the leading term in the large $N$ approximation leads to exact results for the quadratic rephasing invariant average values but for $N=3$ it overestimates the quartic Majorana average values by a factor of $4/3$ and the quartic Dirac average values by a factor of $35/24$.  Hence higher order corrections in $1/N$ are necessary to obtain good approximations for the quartic rephasing invariants when $N=3$.


\section{Conclusion}\label{SecConc}

In this paper we studied analytically the statistical implications of the Haar measure for the rephasing invariants of the PMNS matrix as a function of the number of neutrinos.  After a review of the rephasing invariants and the Haar measure, we introduced the Mellin transform approach to determine the PDFs with the help of the moments.  We calculated the latter from a given parametrization for unitary matrices and showed that under the Haar measure, all PDFs for rephasing invariants of the same type are equivalent.  We then computed the three independent PDFs in terms of the Meijer $G$-functions and studied their physical implications.

We first compared our analytical results with numerical results by generating a large sample of unitary matrices and computing their rephasing invariants.  We also studied the behavior of the PDFs around the origin, showing that they peak at that point, implying that the average values of the absolute values of the rephasing invariants tend to zero as the neutrino number increases.

We then investigated the average values of the absolute values of the rephasing invariants by comparing them with experimental values.  We argued that the $N=3$ case is preferred in our statistical analysis.  However, to take into account all rephasing invariants at the same time, it would be necessary to consider the joint PDF for all rephasing invariants.

With this work, we now have the PDFs for all rephasing invariants under the Haar measure, which appears in the anarchy principle.  The PDF for the light neutrino masses originating from the anarchy principle is also known for arbitrary neutrino number, but it is expressed in terms of a complicated multidimensional integral.  It would be of interest to determine an analytic form for these PDFs, maybe in a large $N$ setting.


\ack{
This work is supported by NSERC.
}


\bibliography{RephasingInv}

\end{document}